\documentclass{rstransa}
\pdfoutput=1
\jname{rsta}
\Journal{Phil. Trans. R. Soc}
\usepackage{booktabs}

\definecolor{dc}{RGB}{0, 143, 0} 

\begin{document}
\title{A New Method for Detecting Solar Atmospheric Gravity Waves}
\author{Daniele Calchetti$^1$, Stuart M. Jefferies$^{2,3}$, Bernhard Fleck$^4$,  Francesco Berrilli$^1$ and Dmitriy V. Shcherbik$^2$}

\address{
$^1$Department of Physics, University of Rome Tor Vergata, I 00133, Italy\\
$^2$Department of Physics and Astronomy, Georgia State University, GA 30303, USA\\
$^3$Institute for Astronomy, University of Hawaii, HI 96768-8288, USA\\
$^{4}$ESA Science and Operations Department, c/o NASA/GSFC, MD 20071, USA}

\subject{Solar System, Observational Astronomy, Astrophysics}

\corres{Daniele Calchetti\\
\email{daniele.calchetti@roma2.infn.it}}

\begin{abstract}
Internal gravity waves have been observed in the Earth’s atmosphere and oceans, on Mars and Jupiter, and in the Sun's atmosphere. 
Despite ample evidence for the existence of propagating gravity waves in the Sun's atmosphere, we still do not have a full understanding of their characteristics and overall role for the dynamics and energetics of the solar atmosphere. 
 Here we present a new approach to study the propagation of gravity waves in the solar atmosphere. It is based on calculating the three-dimensional cross-correlation function between the vertical velocities measured at different heights. We apply this new method to  a time series of co-spatial and co-temporal Doppler images obtained by SOHO/MDI and Hinode/SOT as well as to simulations of upward propagating gravity wave packets. We show some preliminary results and outline future developments.

\end{abstract}
\maketitle
%
\section{Introduction}
Internal gravity waves (IGWs) are waves that are driven by buoyancy forces in a stratified medium. 
They play an important role in mixing material and transporting energy and momentum in planetary atmospheres, oceans, and the radiative interior of solar-type stars. 
In the Earth’s atmosphere IGWs have been observed and modeled for many years, both in the lower atmosphere \cite{eckermann1999,shindell2003} as well as at ionospheric heights \cite{crowley1987}. 
Moreover, IGWs play an important role in the coupling of these two regions. 
In the oceans IGWs power the overturning meridional circulation that affects pollutant disposal, marine productivity and global climate \cite{alford2003}.
Elsewhere in the solar system IGWs have been detected on Mars \cite{seiff1976} and Jupiter \cite{young1997,reuter2007}. 
In fact, the viscous damping of IGWs on Jupiter could explain the enigmatic temperature structure of the planet’s thermosphere, addressing a long-standing problem in planetary physics \cite{reuter2007}. 
In astrophysics, IGWs, randomly excited in space and time by overshooting convection, have been invoked to explain the uniform rotation of the solar core \cite{gough1997} and the rotation profile and the surface Lithium abundance of solar-type stars \cite{charbonnel2005}. 
IGWs have also been proposed as an agent for the mechanical heating of stellar atmospheres and coronae \cite{mihalas1981}. 
More recent studies \cite{vigeesh2017,vigeesh2019,vigeesh2020} show the diagnostic potential of the IGWs for magnetic field in the solar atmosphere at different heights above the solar surface. \\
Due to difficulties associated with directly observing IGWs in the Sun, however, they have not garnered the same level of attention they have in these other disciplines. 
Despite these difficulties, there is observational evidence of short wavelength (less than 6 Mm) IGWs in the Sun’s atmosphere
\cite{deubner1989,deubner1989b,kneer1993,straus1997,rutten2003,straus2008,kneer2011}.  
Here we report on an endeavor to directly measure the IGWs in the Sun’s atmosphere and identify how they couple to atmospheric flows.
\section{Simulations}
\begin{figure}[t]
    \centering
    \includegraphics[width=\textwidth]{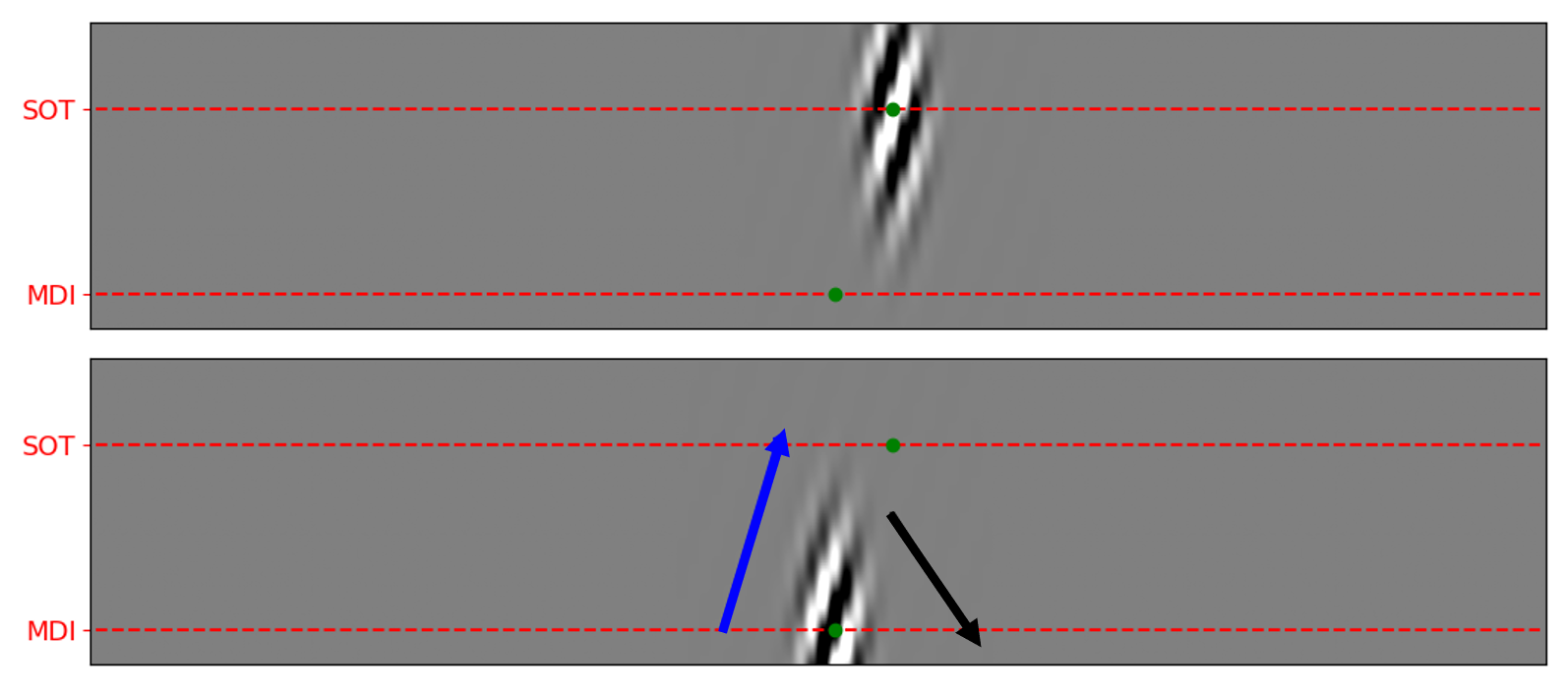}
    \caption{This figure shows the x-z cross-section of a single wave-packet at two times, t$_{0}$ (bottom) and t$_{1}$ (top). At time t$_{1}$ the wave-packet is displaced to the right in x with respect to its location at time t$_{0}$ due the non-zero horizontal group velocity of the wave-packet. In addition, the regions of large amplitude within the wave-packet have moved downward and to the right due to the phase velocities of the waves that comprise the wave-packet. The red dashed lines represent the layers corresponding to SOHO/MDI and Hinode/SOT formation heights. The blue arrow represents the group velocity of the wave-packet, whereas the black arrow represents the phase velocity. Note, in order to clearly see the horizontal displacement of the wave-packet in its journey between the two observing heights, we have displayed the figure with a large aspect ratio. This results in the angle between the arrows showing the directions of the group and phase velocities of the wave-packet appearing to be different from its expected value of 90 degrees.}
    \label{fig:prop}
\end{figure}

Simulations of IGWs can serve as a good (and necessary) bench test to study the propagation of travelling wave-packets. 
The absence of noise and any other disturbances offers fundamental benefits when studying the complex behaviour of wave-packets at different layers in the atmosphere.
We have therefore developed a simple simulation of upward propagating gravity waves in the Sun's atmosphere as a test bench for our analysis. 
In particular, we have created simulated time series of spatially resolved observations of the velocity signals at the two altitudes corresponding to the formation heights of the solar Fraunhofer lines used by the SOHO/MDI \cite{MDI1995} and Hinode/SOT \cite{hinode2007} experiments, with spatial scales and dimensions commensurate with the velocity observations of MDI and SOT (see Section \ref{sec:obs} and Table \ref{tab:data_sim}). 
The parameters of the simulations match exactly the time cadence and spatial resolution of the MDI-SOT dataset. 
This then allows us to directly compare the results of the analyses of our simulated data sets with the analyses of real data we have at hand. \\
We follow the approach of \cite{dornbrack2017} and model
each packet of gravity waves as a traveling wave which produces a perturbation
\begin{equation}
    \zeta (x,y,z,t) = A_{\zeta} \cos(k_x x + k_y y + k_z z - \omega t),
    \label{eq:1}
\end{equation}
where $k_{x,y,z}$ represent the wavenumbers, $\omega$ is the angular frequency and $A_{\zeta}$ is the amplitude of the wave.
We define the angle of propagation of this wave-packet with the equation:
\begin{equation}
    \omega = N \frac{k_h}{\sqrt{k_h^2+k_z^2}} = N \cos\theta
    \label{eq:2}
\end{equation}
where $k_h = \sqrt{k_x^2+k_y^2}$ is the horizontal wavenumber, N is the Brunt-V\"{a}is\"{a}l\"{a} frequency and $\theta$ is the angle between the vertical (z) direction and the group velocity. 
Thus we can calculate the phase and group velocities and the vertical wavenumber for each frequency and horizontal wavenumber:
\begin{equation}
    v_{p(x,y,z)} = \frac{\omega}{k_{(x,y,z)}}
\end{equation}
\begin{equation}
    c_{g(x,y,z)} = \frac{\partial\omega}{\partial k_{(x,y,z)}}
\end{equation}
\begin{equation}
    k_z^2 = k_h^2\left( \frac{N^2}{\omega ^2}-1 \right)
\end{equation}
where $v_{p(x,y,z)}$ and $c_{g(x,y,z)}$ are the components of phase and group velocity respectively. 
For IGWs, the group and phase velocity vectors are perpendicular to each other.
Therefore they have opposite sign in the z-direction (see Figure \ref{fig:prop}). \\
A Gaussian-shaped wave-packet is then completed by defining the amplitude as:
\begin{equation}\label{eq:ampl}
   A_{\zeta}\, \exp\left[ -\left( \frac{(x-x_0-c_{gx}t)^2}{\sigma_x^2} + \frac{(y-y_0-c_{gy}t)^2}{\sigma_y^2} + \frac{(z-z_0-c_{gz}t)^2}{\sigma_z^2}\right)\right]
\end{equation}
where $(x_0, y_0, z_0)$ are the initial positions and $\sigma_x$, $\sigma_y$ and $\sigma_z$ are the physical widths and depth of the wave-packet. \\
Up to this point we have considered an atmosphere without a mean flow, i.e. $U=0$.
We can also assume a time and height independent flow that modifies equation \ref{eq:1}:
\begin{equation}
    \zeta (x,y,z,t) = A_{\zeta} \cos(k_x x + k_y y + k_z z - (\omega + U_x k_x + U_y k_y) t),
\end{equation}
where $(U_x, U_y)$ are the horizontal components of the flow. \\
As IGWs in the Sun's atmosphere are believed to be randomly excited in space and time by convective tongues that penetrate into the stratified atmosphere, the resultant IGW wavefield is random \cite{lighthill1967}.

To mimic this behavior, we use several tens of thousands of wave-packets generated with random parameter values selected from the ranges given in Table  \ref{tab:data_sim}. 
We note that we assume the distribution for each parameter to be uniform over its range and acknowledge that this is a crude approximation. 
We then performed simulations with and without background flows. 
Figure \ref{fig:prop} shows a slice through an example wave-packet observed at two instances in time. 
\section{Observations}\label{sec:obs}
To anchor our simulations we also analyze a set of real (simultaneous) multi-height Dopplergram time series, which were acquired on 20 October 2007 by Hinode/SOT in the Mg b$_2$ line and by SOHO/MDI in its high-resolution mode in the Ni line at 6768 \r{A}. 
Table \ref{tab:data_sim} summarizes the properties of this coordinated dataset. 
The formation height of the MDI Doppler signal is from \cite{fleck2011}, and the mean formation height of the SOT Doppler signal was estimated from the slope of the linear part of the phase difference spectrum between the two lines \cite{straus2009}. 
Other information about this particular data set and the data reduction process can be found in \cite{straus2009}.

\begin{table}[h]
    \centering
    \begin{tabular}{cc}
    \toprule
     Number of pixels along X axis & 256 pix \\
        \midrule
        Number of pixels along Y axis & 256 pix \\
        \midrule
    Number of frames & 735 \\
    \midrule
    Cadence & 60 s \\
        \midrule    
    Duration & 12.25 hours \\
        \midrule
        Pixel scale & 0.61 arcsec/pix \\
        \midrule
        Wavenumber resolution & 0.056 Mm$^{-1}$ \\
        \midrule
        Frequency resolution & 0.023 mHz \\
        \midrule
        MDI formation height \cite{fleck2011} & 125 km \\
        \midrule
        SOT Mg b$_2$ formation height \cite{straus2009} & 720 km \\
    \bottomrule
    \end{tabular}
    \hspace{1cm}
        \begin{tabular}{cc}
    \toprule
        $x_0$ & [0, 255] pix \\
        \midrule
        $y_0$ & [0, 255] pix \\
        \midrule
        $z_0$ & [-100, 100] km \\
        \midrule
        $t_0$ & [0, 734] pix \\
        \midrule
        $\sigma_x$ & [500, 1500] km \\
        \midrule
        $\sigma_y$ & [500, 1500] km \\
        \midrule
        $\sigma_z$ & [100, 500] km \\
        \midrule
        $A_{\zeta}$ & [100, 250] m/s \\
        \midrule
        $k_h$ & [1, 6] Mm$^{-1}$ \\
        \midrule
        $\omega/2\pi$ & [0.5, 3.5] mHz\\
    \bottomrule
    \end{tabular}
    \caption{\textit{Left panel}: Properties of the dataset used in the analysis \cite{straus2009}. \textit{Right panel}: Boundaries of the simulation input parameters. The parameters for the wave-packet simulations have been chosen according to \cite{mihalas1981,mihalas1982,murawski2016}.}
    \label{tab:data_sim}
\end{table}
\section{Analysis}
"Seismic interferometry" shows that the cross-correlation of simultaneous recordings of a random wavefield made at two locations is formally related to the impulse response between those locations \cite{boschi2015}.
Inspired by this and the success of time-distance helioseismology \cite{duvall1993}, we present a new analysis to find the signature of IGWs travelling through the solar atmosphere. 
The method is based on 3D cross-correlation between Dopplergrams at two different heights above the base of the Sun's photosphere. 
\begin{figure}[ht]
    \centering
    \includegraphics[width=.49\textwidth]{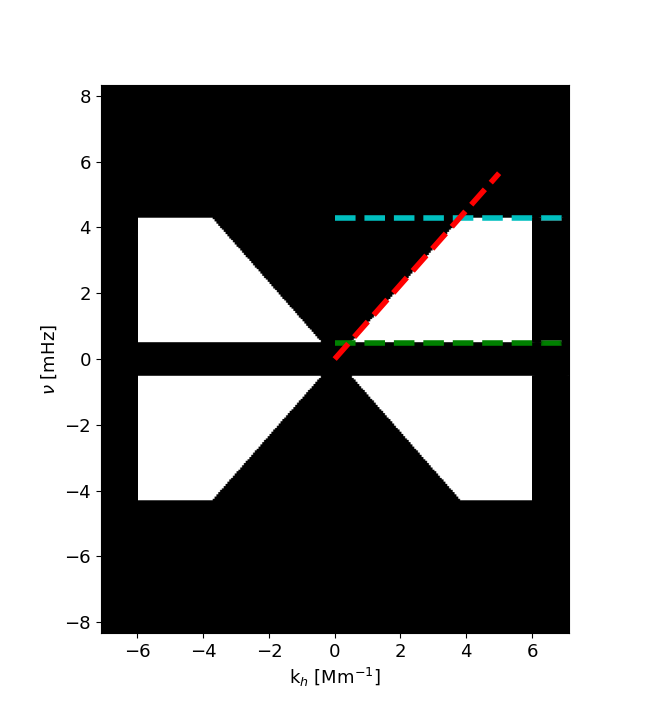}
    \caption{2D section of the subsonic filter used to select the gravity waves regime with 0 points in the center. Red dashed line is the Lamb line, cyan and green horizontal lines are the upper and lower frequency limit. }
    \label{fig:filt}
\end{figure}

The first step in the analysis is the creation of a filter mask in order to select only the gravity waves regime of the power spectrum. 
This corresponds to the region below the Lamb line, defined as $\omega = c_s k_h$, where $c_s$ is the sound speed, and below the Brunt-V\"{a}is\"{a}l\"{a} frequency (which we take as 4 mHz). 
As we will compare the results of our analyses using simulated and real data, we also neglect frequencies below 0.5 mHz and wave numbers above 6 Mm$^{-1}$.  
The former is to reduce the level of convective signal in the real data, the latter is to avoid the observed increased noise in the signal at the higher wavenumbers in the real data (see left panel of Fig. \ref{fig:phMDI-SOT}).
Figure \ref{fig:filt} shows the 3D filter at different wavenumbers and frequencies.
The second step is to modify this filter to select the frequency of the waves we want to analyze. 
We do this by multiplying the filter by an additional Gaussian filter  centered on the frequency of interest with a full-width-at- half-maximum (FWHM) of 0.5 mHz.
The combined filter is then applied to the 3D Fourier transforms of the two data cubes. 
The final step is to compute the cross-spectrum by multiplying one filtered spectrum by the complex conjugate of the other one, and then inverse Fourier transform the cross-spectrum to provide
the 3D cross-correlation function.
\section{Results}
We first look at the phase difference $k_{h}-\omega$ diagram for the real and simulated data as this is historically the tool that has been used to validate the presence of gravity waves in the Sun's atmosphere.
This diagram is generated by computing the phase of the cross-spectrum and then azimuthally averaging over $k_{x}$ and $k_{y}$ to give the temporal variation as a function of $k_{h}$.

\begin{figure}[ht]
    \centering
    \includegraphics[width=.7\textwidth]{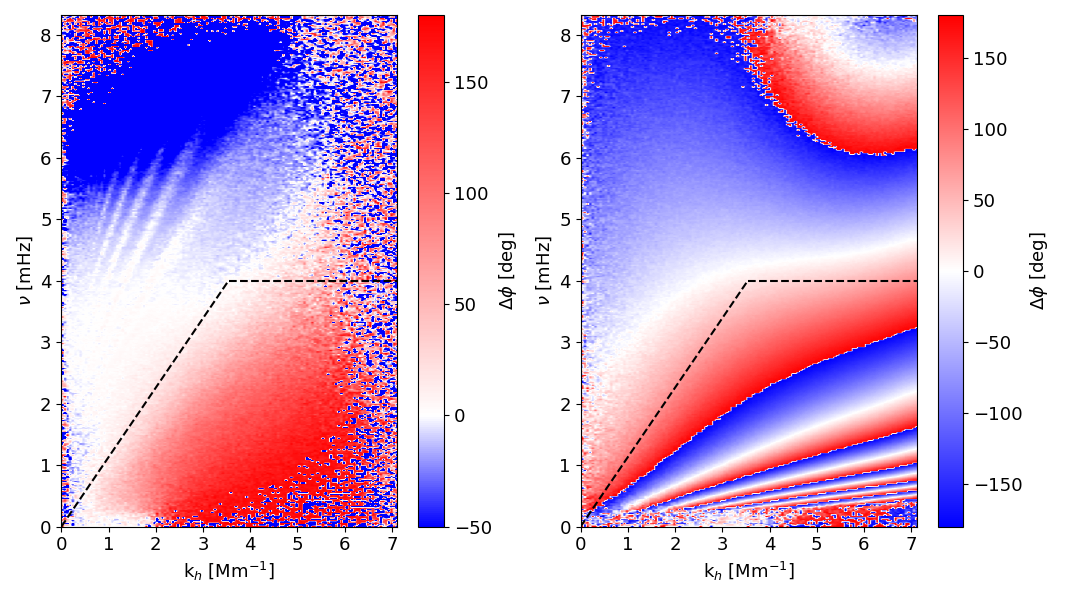}
    \caption{The phase difference $k_{h}- \omega$ diagram for the SOHO/MDI and  Hinode/SOT Doppler observations (\textit{left panel}) and the simulated Doppler data (\textit{right panel}). Both diagrams are computed before the filtering procedure. Black dashed lines represent the Lamb line and the Brunt-V\"as\"ail\"a frequency. Positive angles (red) represent downward phase velocity. The gravity wave signal is visible between 1 Mm$^{-1}$ and 6 Mm$^{-1}$ and below the Lamb line up to about 4 mHz. The phase difference diagram for the simulations shows multiple 360$^\circ$ phase wrappings at low temporal frequencies.
    }
    \label{fig:phMDI-SOT}
\end{figure}

\begin{figure}[ht]
    \centering
    \includegraphics[width=\textwidth]{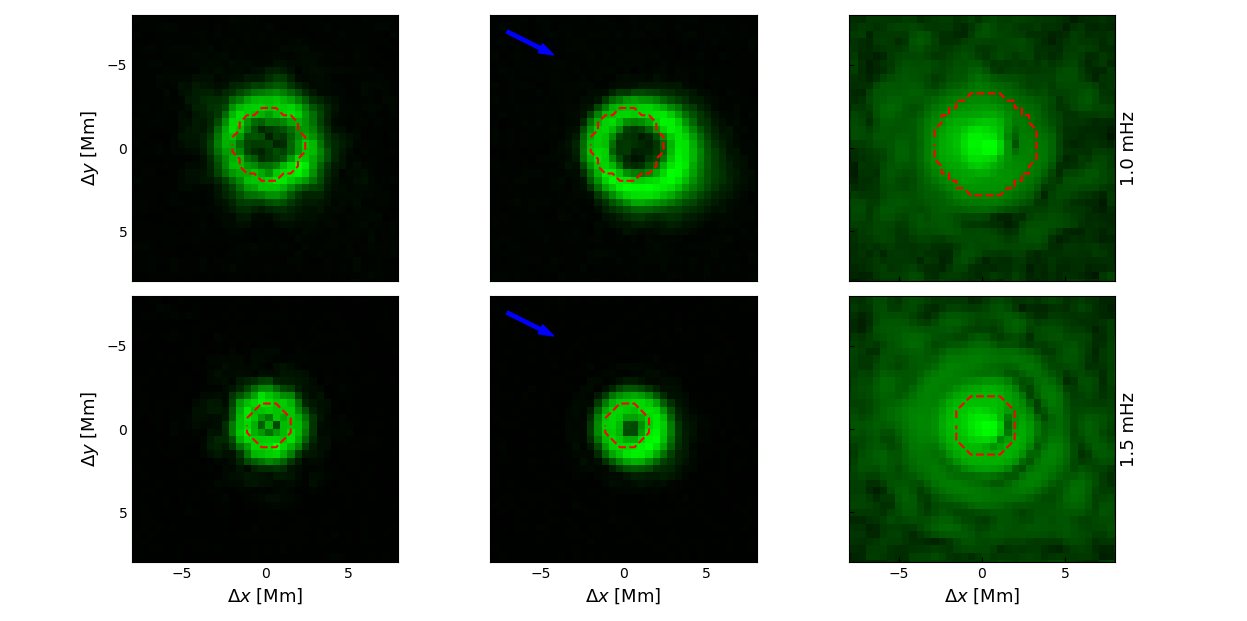}
    \caption{Maximum values along the time axis of the 3D cross-correlation function at 1 mHz (\textit{top row}) and 1.5 mHz (\textit{bottom row}) for different datasets. \textit{Left column}: IGWs simulation without flows. \textit{Central column}: IGWs simulation with flow $U\approx$ 220 m/s. \textit{Right column}: MDI-SOT data cubes. The red circles superimposed on the cross-correlation signals for the simulated data show the expected location of maximum correlation for waves at the central frequency of the frequency filter and with no background flow. The red circles superimposed on the cross-correlation signals for the real data show the expected location for the maximum correlation for a Brunt-V\"as\"ail\"a frequency of 5.5 mHz and a height separation of 600 km. The blue arrows in the images in the central column represent the direction of the background flow.
    The results for the simulated data are shown with linear scaling, whereas the results for the real data are shown with logarithmic scaling. The latter is necessary due to the large correlation near zero lag which comes from non-wave sources (e.g., convection) which are absent in the simulations.}
    \label{fig:2dccf}
\end{figure}

Looking at the the $k_{h} - \omega$ phase difference diagram between the two Doppler signals from the SOHO/MDI and Hinode/SOT data (left panel of Figure~\ref{fig:phMDI-SOT}), we  see the signature of downward phase velocities in the gravity wave regime below the Lamb line. 
This is what we expect for upward propagating gravity waves, whose vertical group and phase velocities have opposite signs. 
The $k_{h}-\omega$ phase difference for the simulated data (right panel of Figure~\ref{fig:phMDI-SOT}) shows the same behavior although with a larger range of phase difference (highlighted by the phase wrapping of the signal).
\\
Next we look at the cross-correlation functions for the real and simulated data. 
In particular, we look at the maximum values of the 3D cross-correlation functions over all temporal lags (Figure~\ref{fig:2dccf}). 
What we see are clear rings at spatial lags that vary with the frequency of the waves under study.  

\begin{figure}[ht]
    \centering
    \includegraphics[width=\textwidth]{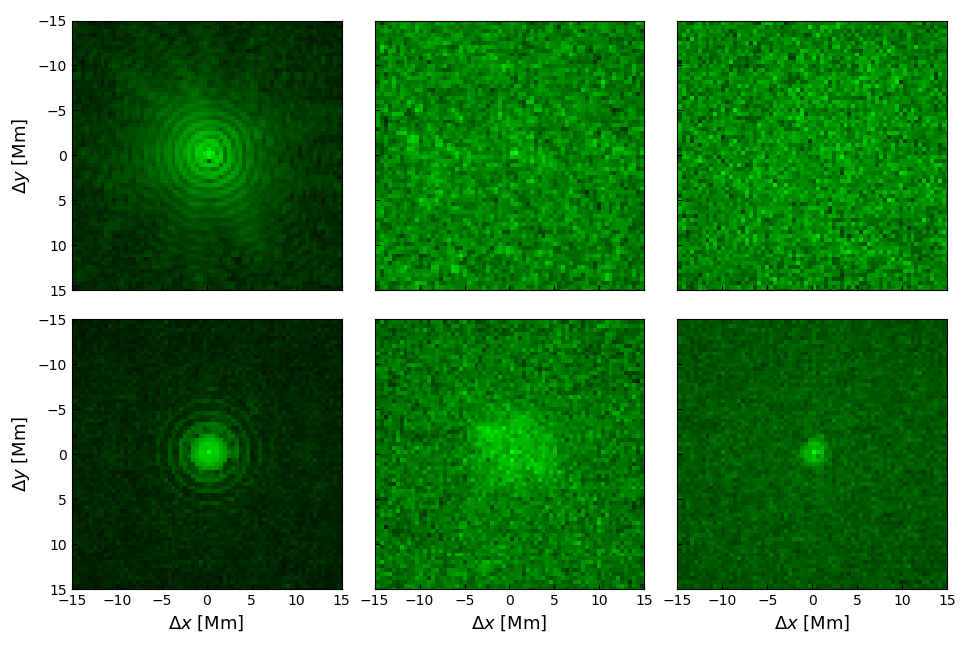}
    \caption {Maximum values along the time axis of the 3D cross-correlation function at 3 mHz for different datasets. All images are shown on a logarithmic scale. \textit{Top-left panel}: MDI-SOT data cubes. \textit{Top-central panel}: shuffled MDI-SOT data cubes. \textit{Top-right panel}: white noise data cubes. \textit{Bottom-left panel}: IGWs simulation without flows. \textit{Bottom-central panel}: shuffled IGWs simulation without flows. \textit{Bottom-right panel}: simulation of running plane waves instead of Gaussian-shaped wavepackets (amplitude in \ref{eq:ampl} set to unity).}
    \label{fig:artifacts}
\end{figure}

To understand this behavior, let's return to Figure \ref{fig:prop} which shows how a wave-packet propagates through the atmosphere. 
The cross-correlation function for the two signals observed at the "SOT" and "MDI" heights in this figure will show a high amplitude at the temporal lag corresponding to the group travel time between the two observing heights and at the spatial lag corresponding to the horizontal distance travelled due to the wave-packet’s horizontal group velocity and the time taken for the wave-packet to travel between the two observing heights. 
If we move to three-dimensions and consider a large number of wave-packets emitted at the same angle (i.e. the wave-packets have the same central frequency) with respect to the vertical, but in all azimuthal directions (i.e. they trace out a cone in 3-D), then we can see that we will observe (2n-1) rings in a cut of the 3-D cross-correlation function at a given time lag, where n is the number of regions of maximum (or minimum) amplitude across the wave-packet. 
If we look at the location of the maximum value in the cross-correlation function over a large number of realizations, we will find that it traces out an annulus whose width is roughly the width of the wave-packet (due to the movement of the peaks due to the phase velocity). 
If we now superimpose on the measured cross-correlation signals the expected locations for the maximum signal according to Equation (\ref{eq:2}) - the red circles in Fig. \ref{fig:2dccf} - it is clear that the signal in the results for the simulation with zero background flow are consistent with the detection of packets of IGWs. Moreover, the real data show the same behavior when the Brunt-V\"as\"ail\"a frequency is set at 5.5 mHz. \\
To further validate that the observed ring structure is due to IGWs and is not an artifact of our filtering method, we performed a number of experiments.
Figure \ref{fig:artifacts} shows the cross-correlation results on different datasets using the filtering procedure at 3 mHz. 
Looking at the figure representing the MDI-SOT dataset and the simulation without flows, we can clearly see rings around the zero-spatial lag point. 
However, when we analyze a white noise data cube, data cubes that have been scrambled in time, and a simulation made of traveling waves where the amplitude of the waves is not limited by equation \ref{eq:ampl}, the results are very different. 
There is no clear ring structure.
This is a clear evidence that the rings visible in the cross-correlations of the real data and in the IGWs simulation, are not artifacts. \\
We now examine how a background flow affects our measured cross-correlation function.
The middle column of Fig. \ref{fig:2dccf} shows the cross-correlation function when there is a constant background flow of~220 m/s. 
What is clear here is that the flow causes the ring in the cross-correlation function to be shifted in the direction of the flow. 
The ring also shows some distortion (which is more evident at higher flow speeds). 
A comparison of the cross-correlation functions at 1 mHz and 1.5 mHz show that cross-correlation function at the lower frequency is influenced more by the flow. 
This is due to the lower frequency wave-packet spending a longer time in the flow (due to its shallower propagation angle and lower velocity): it is therefore more affected by the flow.
\section{Discussion}
Using a simple model of gravity wave excitation and propagation in the solar atmosphere, we have shown that the 3-D cross-correlation function of the time-series of velocity data acquired at two heights in the atmosphere show a unique ring-like structure that varies with the frequency of the waves under study. 

\begin{figure}[h]
    \centering
    \includegraphics[width=.49\textwidth]{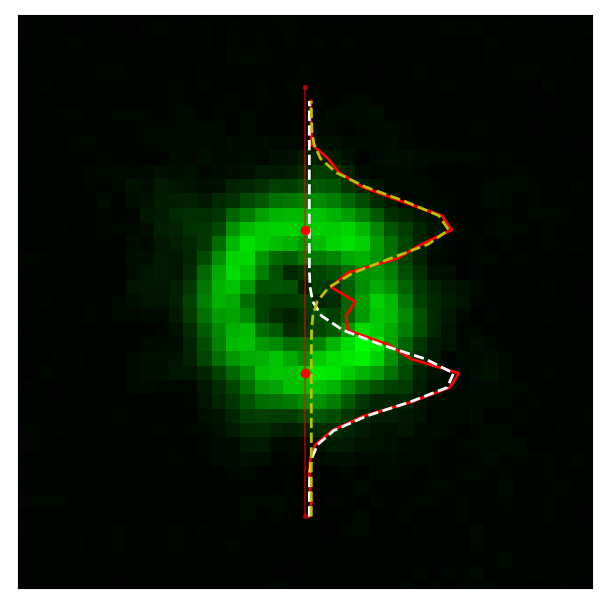}
    \includegraphics[width=.49\textwidth]{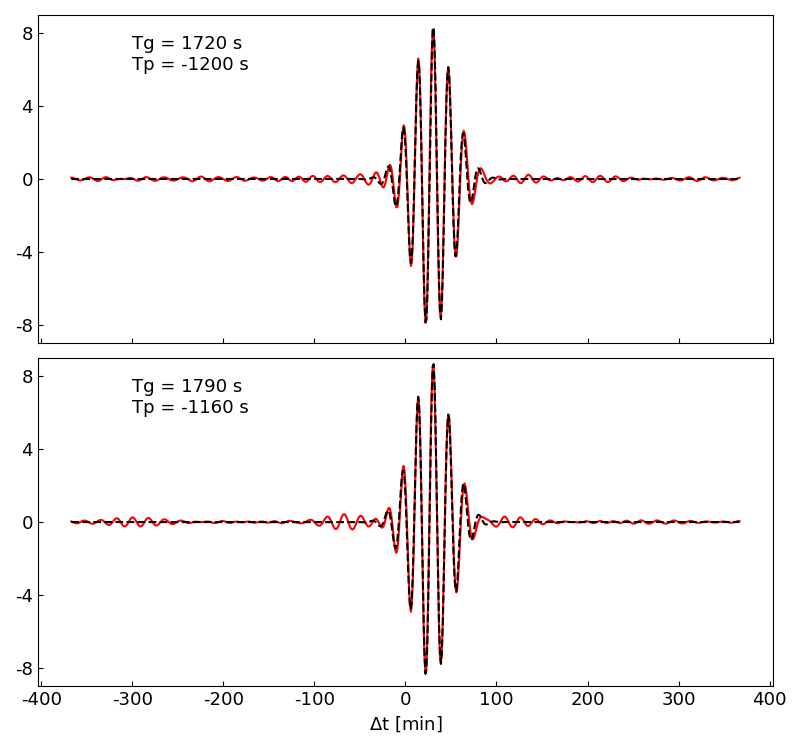}
    \caption{\textit{Left panel}: The spatial distribution of the maximum values along the time axis of the 3D cross-correlation function at 1 mHz for the simulated data without a background flow. The plot inside the figure (red line) represents a vertical cut through the ring of signal. The yellow and white dashed lines are Gaussian fits of the top and bottom sections of the ring. The separation between the two Gaussian fits agrees with the expected diameter of the ring to within 0.1 pixels. 
    \textit{Right panel}: The cross-correlation function as a function of temporal lag at the two positions in the ring in the left plot denoted by the red spots. The data is shown by the red line. The fit to the data  using Equation 1 of \cite{finsterle2004} is shown by the black dashed line. The fit parameters include the group and phase travel times. The values for these quantities, located at the top left corner of each plot, have opposite signs, as expected for IGWs.}
    \label{fig:cut}
\end{figure}
Moreover, we have shown that the characteristics of this ring structure are sensitive to flows that are present between the two observing heights in the solar atmosphere. 
These results suggest that we have uncovered a method to directly measure IGWs  in the solar atmosphere. \\
Measurements of the ring structures at different frequencies and modeling of the temporal variation of the cross-correlation functions in the rings are both straightforward (see Fig. \ref{fig:cut}) and provide a way to infer the properties of the detected gravity wave-packets (e.g, group and phase travel times) and the solar atmosphere (e.g., through the Brunt-V\"as\"ail\"a frequency and sensitivity to flows).
If validated, the results in this paper open the door for using IGWs as a probe of the dynamics in the solar atmosphere (e.g., \cite{Hague2016}). \\
We close with some cautious speculation. Close examination of Fig. \ref{fig:2dccf} provides a tentative hint that the real data may be showing  evidence for a small background flow. However, more research needs to be performed before this conclusion can be drawn. 
Also, there is evidence in the literature for wave reﬂection in the atmosphere and downward propagating wave packets \cite{jefferies1997,rajaguru2013,khomenko2015,jefferies2019}, but our simulation and analysis do not include this yet. 
The effects of flows and wave reflection / boundary layers will be the subject of a future study. 

\dataccess{The SOHO/MDI data used in this study are available from the Joint Science Operations Center (JSOC) at Stanford University (\href{http://jsoc.stanford.edu}{http://jsoc.stanford.edu}) and the Hinode/SOT data from the Hinode Science Data Centre Europe at the University of Oslo, Norway (\href{http://sdc.uio.no/sdc/}{http://sdc.uio.no/sdc/}). 
}
\aucontribute{DC, BF and SJ developed the cross-correlation analysis technique. DC designed and carried out the simulations. DC and DS performed the data analysis. DC and SJ drafted the manuscript. All authors contributed to the discussion and helped polish the manuscript.}
\competing{The author(s) declare that they have no competing interests.}
\funding{
The contributions from SJ and DS to this work were funded under award 1829258 from the National Science Foundation.  DC received travel support from this same award for a trip to Georgia State University. DC was supported by the joint PhD  program in Astronomy, Astrophysics and Space Science between the University of Rome “Tor Vergata”, the Sapienza-University of Rome and the National Institute of Astrophysics (INAF).}

\ack{Hinode is a Japanese mission developed and launched by ISAS/JAXA, collaborating with NAOJ as a domestic partner, NASA and STFC (UK) as international partners. Scientific operation of the Hinode mission is conducted by the Hinode science team organized at ISAS/JAXA. This team mainly consists of scientists from institutes in the partner countries. Support for the post-launch operation is provided by JAXA and NAOJ(Japan), STFC (U.K.), NASA, ESA, and NSC (Norway). SOHO is a mission of international cooperation between ESA and NASA. We thank the referees for constructive comments, which helped to improve this paper. We wish to acknowledge scientific discussions with the Waves in the Lower Solar Atmosphere (WaLSA; \href{https://www.WaLSA.team}{www.WaLSA.team}) team, which is supported by the Research Council of Norway (project no. 262622) and the Royal Society (award no. Hooke18b/SCTM).}
%
%
\bibliography{biblio}
\end{document}